\documentstyle[prb,aps,multicol,epsf]{revtex}

\tighten
\narrowtext

\begin{document}
\draft

\title{Inelastic Scattering Time for Conductance Fluctuations}

\author{I.~L.~Aleiner$^a$ and Ya.~M.~Blanter$^{b}$}
\address{
$^a$ Department of Physics and Astronomy, SUNY at Stony Brook, Stony
Brook, NY 11794\\
$^b$ Department of Applied Physics and DIMES, Delft University of
Technology, Lorentzweg 1, 2628 CJ Delft, The Netherlands}
\date{\today}
\maketitle
\begin{abstract}
We revisit the
problem of inelastic times governing the temperature behavior of the
weak localization correction and mesoscopic fluctuations in one- and
two-dimensional systems. It is shown that, for dephasing by the
electron electron interaction, not only are those times identical but
the scaling functions are also the same.
\end{abstract}


\begin{multicols}{2}

\section{Introduction}

In 1982, Altshuler, Aronov, and Khmelnitsky (AAK) established
\cite{AAK} that electron-electron scattering in metals is
characterized by {\em three} (generally, distinct) time scales. 
These scales are phase-relaxation time $\tau_{\phi}$,
energy-relaxation time $\tau_E$, and out-scattering time
$\tau_e$. The former one is quantum-mechanical and has no classical
analog, while the two latter have a semi-classical interpretation in
terms of Boltzmann equation. The three scales differ in the case
when the energy transferred between electrons in one collision is
small as compared to the temperature of the system $T$.

One can understand the difference between $\tau_E$ and $\tau_e$ by
considering the inelastic collision integral in Boltzmann equation,
\begin{eqnarray}
&&\mbox{St}\left\{
f(\epsilon)
\right\}\! =\! \int\! d\epsilon_1 d\omega K(\omega)
\label{eq1}\\
&&\quad \times
\Big\{\!
- f(\epsilon)\left[1-f(\epsilon-\omega)\right] 
f(\epsilon_1)\left[1-f(\epsilon_1+\omega)\right]
\quad \mbox{(out)}
\nonumber\\
&&\quad \ \ +
\left[1- f(\epsilon)\right] f(\epsilon-\omega)
\left[1-f(\epsilon_1)\right]f(\epsilon_1 + \omega)\Big\},
\quad\mbox{(in)} 
\nonumber
\end{eqnarray}
where $f(\epsilon)$ is the electron distribution function, and the
kernel $K(\omega)$ characterizes matrix elements of the
interaction, with the energy transfer $\omega$. In clean $2D$ and $3D$
systems, $K(\omega)$ is independent on the transmitted energy
$\omega$, $K(\omega)\simeq 1/\epsilon_F$. This results in a Fermi liquid
behavior of the inelastic rate $1/\tau_{in} \simeq {\rm max}(\epsilon,
T)^2/\epsilon_F$. The situation in disordered systems, however, is
different\cite{Schmid,AA79,AALR,AA}, the kernel $K(\omega)$ grows with
the decrease of the transmitted frequency $\omega$,
\begin{equation}
K(\omega) \simeq \frac{1}{|\omega| g(L_{\omega})} \propto \omega^{d/2-2},
\label{eq2}
\end{equation}
where $g(L_\omega)$ is the dimensionless conductance (in units if
$e^2/\pi\hbar$) of the $d$ - dimensional disordered sample of the size
$L_{\omega} = (D/\omega)^{1/2}$, where $D$ is the diffusion constant
of the metallic sample.  For more details on origin of Eq.~(\ref{eq2}),
 see {\em e.g.} Refs.~\onlinecite{Blanter,AAG}. 

Substituting Eq.~(\ref{eq2}) into Eq.~(\ref{eq1}), one estimates
\cite{AA} 
\begin{equation}
\mbox{St}\left\{
f(\epsilon)
\right\}\! \approx - \frac{\delta f(\epsilon)}{\tau_E},
\quad \frac{\hbar}{\tau_E} \simeq \frac{\epsilon^*}{g(L_{\epsilon^*})}
\label{eq3}
\end{equation}
where $\epsilon^*=\mbox{max} (\epsilon, T)$.
Equations (\ref{eq2}) and (\ref{eq3}) are applicable for systems in
the metallic regime, $g(L) \gg 1$. In this regime $\epsilon^* \tau_E
\gg \hbar$, {\em i.e.} quasiparticles are well-defined. 
Notice that,  even though the kernel $K$ is divergent, the energy
relaxation rate (\ref{eq3}) is finite because of the two energy
integrations in Eq.~(\ref{eq1}). Therefore, for the study of the
phenomena governed by the Boltzmann equation, the infrared divergence
of the matrix elements (\ref{eq2}) does not cause any problems.
These phenomena include, for instance, electron distribution function
measured via tunneling spectroscopy \cite{Devoret} or crossover from
$1/3$ to $\sqrt{3}/4$ shot noise in metallic wires \cite{Noise}.

It is not the end of the story, though.  If we estimate only one
(``out'') term from the collision integral (\ref{eq1}), we encounter
an infrared divergence in two- and one-dimensional cases,
\begin{equation}
\mbox{St}_{out}\left\{ f(\epsilon) \right\}\! \approx - \frac{\delta
f(\epsilon)}{\tau_e}, \quad \frac{\hbar}{\tau_e} = \frac{T}{g(L_T)}
\int_{\omega^*}^T \frac{d \omega}{\omega}
\left(\frac{T}{\omega}\right)^{\frac{2-d}{2}},
\label{eq4}
\end{equation}
where $\omega^*$ is the low energy cut-off to be found,
and $L_T = \sqrt{D/T}$ is the temperature length.
(The same result may be obtained from the calculation of the first
loop correction to the self-energy\cite{AALR}.) This divergence of
only one contribution to the collision integral is a simple
consequence of the fact that each term in collision integral is not a
gauge invariant quantity, and only both terms taken together have a
physical meaning (\ref{eq3}), which is not cut-off dependent. One can
argue, however, that $\tau_e$ has its own observable consequences for
{\em the quantum interference processes}. Indeed, naive argument is
that the ``out'' processes completely suppress the interference,
whereas ''in'' processes are incoherent. Inclusion of some of the
higher order processes\cite{AALR,Blanter} cures the divergence and
makes the expression for $1/\tau_e$ finite. One may naively expect
that $\tau_e$ found from such procedure is, indeed, responsible for
the temperature behavior of quantum corrections.   

AAK showed\cite{AAK} that it is not correct for the temperature
behavior of weak localization correction, because the inelastic
excitations with energy transfer smaller than decoherence rate itself
do not suppress this correction, see Section 2.2.2 of
Ref.~\onlinecite{AAG} and our Section~\ref{sec:2} for the
corresponding physical argument. This leads to the infrared cut-off
$\omega^* \simeq 1/\tau_{\phi}$ in Eq.~(\ref{eq4}) and to the
self-consistency equation for the dephasing rate,
\begin{equation}
\frac{\hbar}{\tau_{\phi}} \simeq \frac{T}{g(L_\phi)},
\quad L_\phi = \sqrt{D\tau_\phi}.
\label{eq5}
\end{equation}
 
However, there is a prejudice, see {\em e.g.} Ref.~\onlinecite{Blanter}, 
that the inelastic time governing the magnitude of the
conductance fluctuation is given by $\tau_e \ll \tau_\phi$, so that 
$\tau_e$ has its own observable effect.

In this paper we revisit this problem. We will show that the inelastic
rate governing the mesoscopic fluctuations is precisely the same as
for the weak localization, see Eq.~(\ref{eq5}). Moreover, the scaling
functions governing the magnetic field and the temperature behavior of
conductance fluctuations are found to be identical to their weak
localization counterparts, see Sections~\ref{sec:3}, \ref{sec:4}.

The remainder of the paper is arranged as follows.
Section \ref{sec:2} is devoted to the qualitative discussion
of the role of the effect of the real electron-hole pair excitations
on the weak localization and mesoscopic conductance fluctuations.
The main point of this Section is to explain why the singlet
excitations with transmitted frequency smaller than $1/\tau_\phi$
affect neither weak localization nor mesoscopic fluctuations.
In Section~\ref{sec:3} we explicitly calculate the effect of
interactions on mesoscopic fluctuations of conductance in one
dimension, using the same approach as AAK \cite{AAK}. We will also
identify the diagrammatic contributions which are missed in the
arguments for the role of $\tau_e \ll \tau_\phi$ in the conductance
fluctuations. Section~\ref{sec:4} generalizes the calculation to two
dimensions. Our findings are summarized in Conclusions.

\section{Qualitative discussion}
\label{sec:2}

The purpose of this Section is to explain interference processes,
taking into account possibility of excitations of real electron-hole
pairs, see also
Ref.~\onlinecite{Stern}.
 For the weak localization correction, similar arguments were
used in Ref.~\onlinecite{AAG}.   

A qualitative physical interpretation of quantum corrections
is usually based on the following arguments, see {\em e.g.}
Ref.~\onlinecite{ALee}. Consider an electron diffusing in a good
conductor, $p_Fl \gg \hbar$. Probability $w$ for the electron to
reach, say, point $i$ starting from point $f$, see Fig.~\ref{Fig1}a,
\ref{Fig2}a, can be obtained by first finding the semiclassical
amplitudes $A_\alpha$ for different paths connecting the points, and
then, calculating the absolute value of their sum,
\begin{equation}
w = \left|\sum_\alpha A_\alpha \right|^2 = \sum_\alpha
\left|A_\alpha\right|^2 + \sum_{\alpha\neq\beta} A_\alpha A_\beta^\ast.
\label{eq:2.1}
\end{equation}
The first term in Eq.~(\ref{eq:2.1}) is nothing but the sum of the
classical probabilities of the different paths, and it may be found
from the classical Boltzmann equation. The second term is the
quantum mechanical interference of the different paths. 
In what follows, we will discuss the contribution of this
term to transport and how it is affected by the electron-electron
interaction.

\subsection{ Weak localization correction} 
\label{wl}
For generic pairs $\alpha, \beta$, the product $A_\alpha A_\beta^\ast$
oscillates as the function of impurity configurations, see
Fig.~\ref{Fig2}a.  This is because the lengths of paths $\alpha$ and
$\beta$ are substantially different. As the result, contribution
of such paths is not relevant for disorder averaged quantities but 
contributes to the mesoscopic fluctuations of the conductance.

\begin{figure}
\narrowtext
\epsfxsize=0.8\hsize
\centerline{\epsffile{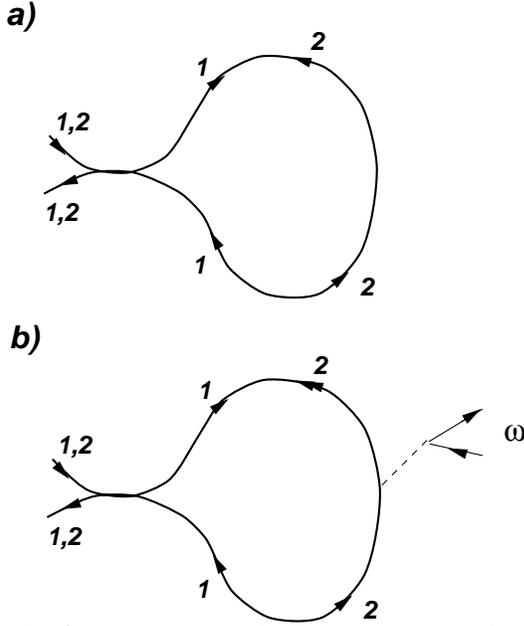}}
\caption{a) Example of classical paths between points $i$ and $f$
contributing to the weak localization; b) The same paths with
the excitation of one electron-hole pair with energy $\omega$.}
\label{Fig1}
\end{figure}

There are pairs of paths, however, which preserve the same phase, with
the change of the disorder configuration. An example of such
paths is shown in Fig.~\ref{Fig1}a. These paths almost coincide
everywhere except the loop segment $BEB$ (see Fig.~\ref{Fig1}a) which
is traversed by trajectories $1$ and $2$ in the opposite
directions. In the absence of the magnetic field and spin-orbit
interactions, the phases of the trajectories $1$ and $2$ are equal.
Therefore, the contribution of these paths to the probability $w$
becomes
\begin{equation}
\left|A_1 + A_2\right|^2 = \left|A_1\right|^2 + \left|A_2\right|^2
+2 {\rm Re} A_1A_2^\ast = 4 \left|A_1\right|^2,
\label{eq:2.2}
\end{equation}
{\em i.e.} twice larger than the classical probability. Thus, in order
to evaluate the weak localization  correction to the conductivity, one
has to determine the classical probability to find such a
self-intersecting trajectory.  

Let us now consider the main effect of electron-electron interactions
on the weak localization --- excitation of soft electron-hole pairs.
We consider processes involving either one excitation (probability
$P_1$) or no excitations (probability $P_0 = 1 - P_1$), see
Fig.~\ref{Fig1}b. Allowing for the excitation of an electron-hole
pair, one obtains 
\begin{equation}
A_\alpha \to A_\alpha^{0} + A_{\alpha}^{1},
\label{eq:2.3}
\end{equation}
where the superscripts $0$ and $1$ correspond to the amplitudes
involving emission of no electron-hole pairs or one electron-hole pair 
respectively.

Because the states with different number of excitations are orthogonal
to each other, we obtain, instead of Eq.~(\ref{eq:2.2}),
\begin{eqnarray}
&&\left|A_1^0 +A_1^1 + A_2^0 + A_2^1\right|^2 = 
\left|A_1^0\right|^2 + \left|A_2^0\right|^2+
\left|A_1^1\right|^2 + \left|A_2^1\right|^2
\nonumber\\
&& 
+2 {\rm Re} A_1^0[A_2^0]^\ast
+2 {\rm Re} A_1^1[A_2^1]^\ast,
\label{eq:2.4}
\end{eqnarray}
where the last two terms correspond to the interference correction.
It is important to emphasize that the interference persists even if
the final state contains an electron-hole excitation (last term).

We now notice that the emission of a soft electron-hole pair
does not alter the geometrical form of the trajectory, thus, it does
not change the classical probability corresponding to path
$\alpha$. As the result, we have   
\begin{equation}
|A_\alpha^0|^2= P_0 |A_\alpha|^2,\quad 
|A_\alpha^1|^2= P_1 |A_\alpha|^2,
\label{eq:2.5}
\end{equation}
where amplitudes without superscript correspond to those in the
absence of the interaction. What the emission of the electron-hole
pair may change, however, is the phase of the quantum amplitude.

Indeed, denote the point of emission of electron-hole pair of
energy $\omega$ on a classical trajectory by $t_\alpha^{em}$ -- time
it takes for the electron moving along the trajectory $\alpha$ with
energy $\epsilon_F$ to reach the emission point, 
see Fig.~\ref{Fig1}. Denote the total time along the path $\alpha$ as 
$t_\alpha$. Than, the electron moves time $t_\alpha^{em}$ with the energy 
$\epsilon$ and time $t_\alpha - t_\alpha^{em}$ with the energy 
$\epsilon -\omega$.  As the result, the geometrical phase, accumulated
by electron, changes as
\[
{\rm arg}A_\alpha^1={\rm arg}A_\alpha^0 - \omega 
\left(t_\alpha - t_\alpha^{em}\right).
\]
Thus,
\begin{eqnarray}
&&
A_\alpha^0[A_\beta^0]^{\ast}=P_0A_\alpha A_\beta^{\ast},
\nonumber\\
&&
A_\alpha^1[A_\beta^1]^{\ast}=P_1 A_\alpha A_\beta^{\ast}
e^{i\omega (t_\beta -t_\alpha - t_\beta^{em}+ t_\alpha^{em})}
.
\label{eq:2.8}
\end{eqnarray}

Substituting Eqs.~(\ref{eq:2.5}) and (\ref{eq:2.8})
into Eq.~(\ref{eq:2.4}), we obtain [instead of Eq.~(\ref{eq:2.2})] 
for paths contributing to the weak localization correction, 
\begin{eqnarray}
&&\left|A_1^0 +A_1^1 + A_2^0 + A_2^1\right|^2
\nonumber\\
&&\quad
=
2 \left|A_1\right|^2 + 2 \left|A_1\right|^2 
\left[P_0 + P_1 \cos\omega(t_1^{em}-t_2^{em})
\right]
.
\label{eq:2.9}
\end{eqnarray}

The last term in Eq.~(\ref{eq:2.9}) describes the effect of the
excitation of an electron-hole pair in the system on the weak
localization correction. One can readily see that not each inelastic
process destroys the interference. For instance, for $\omega \to 0$,
Eq.~(\ref{eq:2.9}) reproduces Eq.~(\ref{eq:2.2}) exactly!! 
On the other hand, the time $t_\alpha^{em}$ is shorter than
$\tau_\phi$. Thus, we may conclude that inelastic processes with
energy transfer $\omega \lesssim 1/\tau_\phi$ do not destroy the
interference, which gives the physical reason for the low energy
cut-off  $\omega^* \simeq 1/\tau_{\phi}$ in Eq.~(\ref{eq4}).

\subsection{Mesoscopic conductance fluctuations.}
\label{mf}

\begin{figure}
\narrowtext
\epsfxsize=0.8\hsize
\centerline{\epsffile{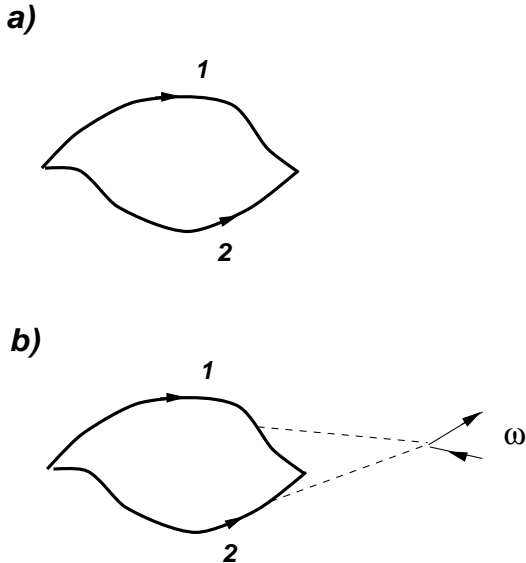}}
\caption{a) Example of classical paths between points $i$ and $f$
contributing to the mesoscopic conductance fluctuations; 
b) The same paths with the excitation of one electron-hole pair with
energy $\omega$.} 
\label{Fig2}
\end{figure}

{\em Effect of inelastic processes}. The arguments of the previous
subsection are easily generalized for the effect of inelastic
processes on mesoscopic conductance fluctuations. We can still talk
about a pair of two paths, but now we will take those paths to be
generic, see Fig.~\ref{Fig2}. The interference contribution from those
paths,  
\begin{equation}
\delta G \sim 2 {\rm Re} A_1A_2^\ast,
\label{eq:2.10}
\end{equation}
does not affect the average conductance because of random phases of
those amplitudes, but it gives rise to the mesoscopic fluctuations of
the conductance,
\begin{equation}
\langle \delta G^2 \rangle  \sim  2 \langle |A_1|^2 \rangle \langle
|A_2|^2 \rangle. 
\label{eq:2.11}
\end{equation}

Let us now consider the effect of the excitation of an electron-hole
pair of energy $\omega$. To do so, we use the qualitative argument of
previous subsection [starting from Eq.~(\ref{eq:2.3})] and substitute
Eq.~(\ref{eq:2.8}) into Eq.~(\ref{eq:2.10}). It yields
\begin{equation}
\delta G \sim 2 {\rm Re} \left[
A_1A_2^\ast (P_0 + P_1e^{i\omega (t_2 -t_1 - t_2^{em}+ t_1^{em})} )
\right]. 
\label{eq:2.12}
\end{equation}

Once again, we arrive to the conclusion that the excitations of
frequencies smaller than the inverse times to traverse the
trajectories, $1/t_{1,2}$, do not change the interference
correction. Similarly to the weak localization the lengths of paths
are limited by $\tau_\phi$.  Thus, we may conclude that inelastic
processes with energy transfer $\omega \lesssim 1/\tau_\phi$ do not
affect mesoscopic fluctuations, which gives the physical reason for
the low energy cut-off $\omega^* \simeq 1/\tau_{\phi}$ in
Eq.~(\ref{eq4}). Thus, inelastic time entering the weak localization
and mesoscopic fluctuations should be approximately the same. The
exact equality of those times will be proven in the next Section by a
direct calculation, however, this result is definitely model
dependent. Namely, it implies that the contribution of the
quasi-static fluctuations in the systems does not overwhelm  the role
of the inelastic processes, and we discuss such fluctuations now.

{\em Effect of quasi-static fluctuations}. In the linear response
theory, a many-body system in its stationary state is excited at some
time $t_1$ and than the behavior of some observable quantity is
studied at times $t > t_1$. If the temperature is finite, the initial
stationary state of the system can be not only its ground state $E_0$, 
but also any of many-body eigenstates, $E_\alpha$; the probability
that the system is initially in such a state is $\propto
e^{-E_{\alpha}/T}$. If there were no interaction, it would result only
in the thermal average of the mesoscopic fluctuations. However,
electron-electron interaction leads to the effective  dependence of
the disordered potential for electrons. The simplest, and the most
effective example of this mechanism is the dependence of the Hartree
potential of the electrons on the electron configuration. Since the
measurable conductance is the result of the large number of
measurements, each time the initial state may be different. 

In principle, one could expect that the averaging over different
configurations of the self-consistent potential may lead to an effect
stronger than the excitations of the electron-hole pairs. This is
possible, when there is an additional slow degree of freedom
such as magnetic impurities\cite{Falko}, moving defects\cite{ImryFS},
or slow fluctuations of the gauge field\cite{Narozhny}. 
However, this is not the case for the Coulomb electron-electron
interaction, as we explain below.  

To find the magnitude of the effect, we first have to estimate
the characteristic value of possible fluctuations, then evaluate the
effect of such fluctuations on the mesoscopic fluctuations of
conductance, and then compare this effect with effect of $\tau_\phi$ 
coming from the inelastic processes. According to Nyquist noise
formula, the amplitude of the electric field $\delta E(L)$ fluctuating
on the spatial scale $L$ is given by
\begin{equation}
\delta E^2(L) \simeq \frac{T}{\sigma_d}\frac{\bar{\omega}}{L^d},
\label{eq:Nyquist}
\end{equation}
where $\sigma_d$ is the conductivity of the system in $d$ dimensions,
and $\bar{\omega} \ll 1/\tau_\phi$ is the high-energy cut-off above
which fluctuations can not be considered as quasistatic.

To have a strong effect on mesoscopic fluctuations, the electric field
should change significantly the wavefunctions of the electrons
on the scale $L_{\phi}$, which translates into the condition\cite{LK}
\begin{equation}
e|\delta E|L_\phi \gtrsim \frac{\hbar D}{L_{\phi}^2} = 
\frac{\hbar}{\tau_{\phi}}.
\label{eq:condition}
\end{equation} 

On the other hand, we estimate from Eq.~(\ref{eq:Nyquist}),
\[
e^2|\delta E(L_{\phi})|^2 L_\phi^2 =T\bar{\omega}
\frac{e^2 L_{\phi}^{2-d}}{\sigma_d} = 
\frac{T\hbar \bar{\omega}}{g(L_{\phi})},
\]
where $g(L)$ is the dimensionless conductance on the linear scale $L$.
Taking into account Eq.~(\ref{eq5}) and the condition
$\bar{\omega}\tau_\phi \ll 1$ we conclude that for the dephasing by
the Coulomb interaction the condition (\ref{eq:condition}) can be
never satisfied, and therefore the quasistatic fluctuations are
negligible in comparison with the inelastic processes.  We reiterate
that this result does not hold for the scattering on the collective
modes, which have peak in their spectral density on frequencies much
smaller than $1/\tau_\phi$.

\section{Conductance fluctuations in quasi-one-dimensional systems} 
\label{sec:3}

In this Section, we consider a quasi-one-dimensional wire of length
$L$ and the number of transverse channels $N_{\perp}$. The static
conductance of the wire $G$ is expressed through the non-local
conductivity $\sigma (x_1, x_2)$ as follows, 
\begin{equation} \label{cond0}
G = \frac{1}{L^2} \int dx_1 dx_2 \sigma_{xx} (x_1, x_2),
\end{equation}
where $x_1$ and $x_2$ label the coordinates along the wire. To
simplify the expressions, we disregard first inelastic processes
and include them later on. We express the symmetric part of the
conductivity in terms of Green's functions and substitute it in
Eq.~(\ref{cond0}). We find\cite{LSF,BarangerStone},  
\begin{equation}
G= \int \frac{d\bbox{r}_1 d\bbox{r}_2}{L^2}\int
\frac{d\epsilon}{\pi} \frac{d f}{d\epsilon} 
\hat j_{x_1} G^R (\bbox{r}_1, \bbox{r}_2; \epsilon) j_{x_2} 
G^A (\bbox{r}_2, \bbox{r}_1; \epsilon),
\label{GGF}
\end{equation}
where the integration is performed over all the sample,
the spin degeneracy is taken into account, $f$ is the
Fermi distribution function,  and the current operator $\hat j_x$ is
defined as follows, 
\begin{displaymath}
g_1 \hat j_x g_2 = \frac{ie}{2m} ( g_2 \partial_x g_1 - g_1 \partial_x
g_2 ).    
\end{displaymath}
For the rest of the article, we employ the system of units with $\hbar
= 1$, and restore $\hbar$ in the final results.

\begin{figure}
\narrowtext
\epsfxsize=0.9\hsize
\centerline{\epsffile{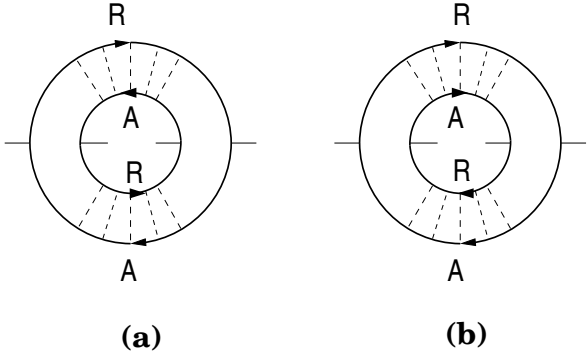}}
\vspace{0.4cm}
\caption{Diagrams with two diffusons (a) and two cooperons (b)
contributing to the conductance fluctuations. Dashed lines represent
impurity scattering, interaction is not yet taken into account.} 
\label{Fig3}
\end{figure}

In the following, we consider only high-temperature limit, $T \gg
D/L^2$, because it is the only case when the inelastic processes
(rather than sample size, $L$)  are controlling the magnitude of the
fluctuations. In this case the main contribution to the conductance
fluctuations is given by two-diffuson and two-cooperon diagrams,
Fig.~\ref{Fig3}. The resulting correlation function for the
conductance fluctuations at different magnetic fields $H_1,\ H_2$ is
expressed in the time domain as 
\begin{eqnarray} \label{cond3}
&& \overline{ \delta G(H_1) \delta G(H_2)} 
=  \frac{(2e^2 D)^2}{3\pi T L^4}\int dx_1 dx_2
\int dt \\
&&\quad\times 
\left[\left|{\cal P}_D^{12} (x_1, x_2,t)\right|^2+
\left|{\cal P}_C^{12} (x_1, x_2; t)\right|^2
\right],
\nonumber
\end{eqnarray}
where the overbar  stands for the disorder averaging.
Deriving Eq.~(\ref{cond3}), one makes use of the approximation
\[
\int \frac{d\epsilon_1}{2\pi} \frac{d\epsilon_2}{2\pi}
\partial_{\epsilon_1} f \partial_{\epsilon_2} f e^{i(\epsilon_1 -
\epsilon_2)(t - t')} \approx \frac{1}{12\pi T} \delta(t - t'), 
\]
justified at time scale larger than $1/T$.

Semiclassical retarded diffuson and cooperon propagators entering into
Eq.~\ref{cond3}) are solutions of the equations
\begin{equation} \label{difnoint}
 \left(  \partial_{t}  - D 
\partial^2_x 
+ \left\{\matrix{\frac{1}{\tau_D^{12}} \cr
\frac{1}{\tau_C^{12}}
}\right\}\right) 
\left\{\matrix{
{\cal P}^{12}_D (x, x'; t)\cr
{\cal P}^{12}_C(x, x'; t)}\right\}
 = \delta (x - x') \delta (t),
\end{equation}
where $D$ is the diffusion coefficient, and the symmetry breaking
parameters $\tau^{12}_{C,D}$ are defined as (see
Ref. \onlinecite{AAG})
\begin{equation}
\frac{1}{\tau_{D}^{12}} = 
\frac{e^2 a^2 D (H_1-H_2)^2}{12\hbar^2 c^2},
\quad
 \frac{1}{\tau_{C}^{12}} = \frac{e^2 a^2 D (H_1 + H_2)^2}{12\hbar^2
c^2}, 
\label{times}
\end{equation}
with $a$ being the transverse dimension of the sample. It is worth
mentioning that the numerical coefficient here is geometry dependent. 

So far, we merely followed a standard avenue (see {\em e.g.}
Ref. \onlinecite{LSF}). Now we are prepared to introduce
electron-electron interactions. On the language of diagrams, we must
add to Fig.~\ref{Fig3} all of the possible interaction lines.  Since inner
and outer rings represent the measurement at significantly different
times, the interaction lines {\em do not} connect these two rings, and
only may be drawn within the same ring, connecting $G^R$ with $G^R$,
$G^R$ with $G^A$, and $G^A$ with $G^A$ for the same impurity
configuration.  Following Ref.~\onlinecite{AAK}, these lines are
conveniently represented by external time-dependent random fields,
$\varphi^{\alpha} (x,t)$, where the index $\alpha$ assumes values
$\alpha = 1$ (outer ring) and $\alpha = 2$ (inner ring). These fields
are assumed to be Gaussian distributed with zero average. The
correlation function is described by the Keldysh component of the
propagator of the screened Coulomb interaction,
\begin{eqnarray} \label{fdt1}  
& &  \langle \varphi^{\alpha} (x,t) \varphi^{\beta} (x',t') \rangle
\nonumber \\
& = & \delta_{\alpha\beta} \delta (t - t') \frac{2T}{D\nu_1} \int
\frac{dq}{2\pi} \frac{1}{q^2} e^{iq(x-x')}, 
\end{eqnarray}
where $\nu_1$ is the thermodynamic density of states per unit length.
Equation (\ref{fdt1}) is nothing but a space-time version of 
\begin{displaymath}
 \langle \varphi \varphi \rangle (q,\omega) = - {\rm Im}\ 
\frac{2T}{\omega} \ \frac{Dq^2 - i\omega}{Dq^2\nu_1},
\end{displaymath}
and we assumed $T \gtrsim \omega$. 
This assumption is justified, because the main contribution to
the dephasing rate is coming from the energy transfer $\omega$
much smaller than $T$. (The diagrams explicitly showing cancellation
of all the processes with $\omega > T$ can be found, {\em e.g.}, in
Refs.~\onlinecite{LSF,Blanter}.) Because we also disregard all
effects due to finite size of the sample, this implies the
following hierarchy of energy scales,
\begin{equation} \label{regime}
T \gg \tau_{\phi}^{-1} \gg E_c \equiv D/L^2
\end{equation}
(here $\tau_{\phi}$ stands not only for the phase-relaxation time, but
for all time scales due to electron-electron scattering). In the
following, we assume that the conditions (\ref{regime}) are 
satisfied. 

\begin{figure}
\narrowtext
\epsfxsize=0.9\hsize
\centerline{\epsffile{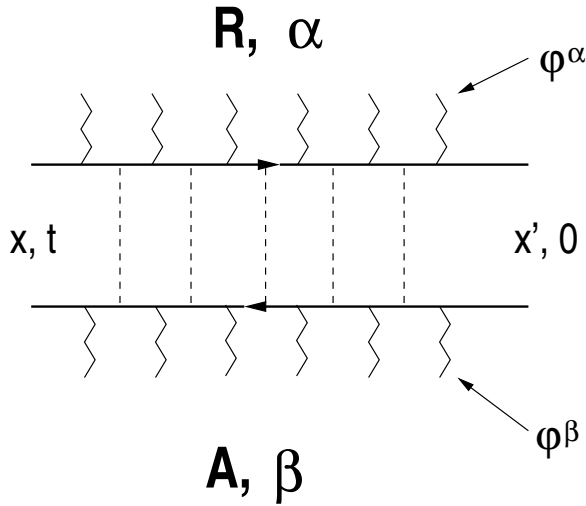}}
\vspace{0.4cm}
\caption{CF diffuson ${\cal P}_D^{\alpha\beta}$. Zigzag lines
represent random fields $\varphi^{\alpha,\beta}$.} 
\label{Fig4}
\end{figure}

The factor $\delta_{\alpha\beta}$ in the right-hand side of
Eq.~(\ref{fdt1}) explicitly indicates that the fields attached to
outer and inner rings of the diagram Fig.~\ref{Fig4} are
uncorrelated, {\em i.e.} no interaction lines, indeed, can be drawn
between the rings. The momentum integral in Eq. (\ref{fdt1}) diverges,
but our final result will contain well-defined differences of
integrals of this type.

Introduction of the fluctuating fields modifies the equations for the
diffuson and cooperon (\ref{difnoint}), see Fig.~\ref{Fig4} and
Ref.~\onlinecite{AAK}, which now become the functionals of the
fluctuating fields, 
\begin{eqnarray} \label{difint}
 &&\left[  \partial_{t}  - D 
\partial^2_x + i \left(\varphi^{\alpha} (x, t) -
\varphi^{\beta} (x, t)\right)
+ \left\{\matrix{\frac{1}{\tau_D^{\alpha\beta}} \cr
\frac{1}{\tau_C^{\alpha\beta}}
}\right\}\right]\nonumber\\
&& \quad \times 
\left\{\matrix{
{\cal P}^{\alpha\beta}_D (x, x'; t; \left\{\varphi^{\alpha} (x, t),
\varphi^{\beta}(x, t)
\right\})\cr
{\cal P}^{\alpha\beta}_C(x, x'; t;
\left\{\varphi^{\alpha} (x, t),
\varphi^{\beta}(x, t)
\right\}
)}\right\} \nonumber \\
& & \quad \quad = \delta (x - x') \delta (t). 
\end{eqnarray}

The correlation function of conductances  is given
by the equation similar to Eq.~(\ref{cond3}), but all the interaction
lines in Eq.~(\ref{difint}) are connected by the propagator
(\ref{fdt1})
\begin{eqnarray} \label{cond4}
&& \overline{ \delta G(H_1) \delta G(H_2)} 
=  \frac{(2e^2 D)^2}{3\pi T L^4}\int dx_1 dx_2
\int dt \\
&&\quad\times 
\left[\langle\left|{\cal P}_D^{12} (x_1, x_2,t)\right|^2
\rangle_{\varphi}
+\langle
\left| {\cal P}_C^{12} (x_1, x_2; t)\right|^2\rangle_{\varphi}
\right],
\nonumber
\end{eqnarray}
where $\langle\dots\rangle_{\varphi}$ stand for the averaging over
the fluctuating field $\varphi^{1,2}$.

\begin{figure}
\narrowtext
\epsfxsize=0.9\hsize
\centerline{\epsffile{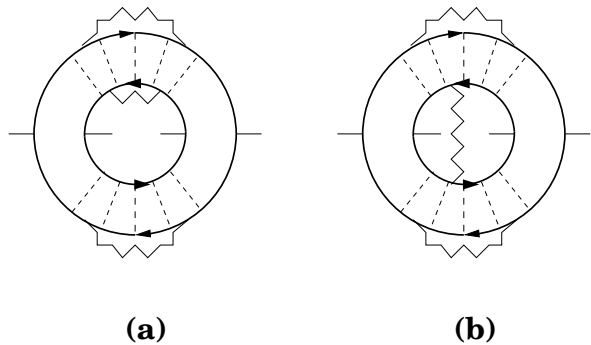}}
\vspace{0.4cm}
\caption{Examples of diagrams with interaction (shown as zigzag lines)
contributing to conductance fluctuations. The diagram (a) is reduced
to the CF diffuson, while the diagram (b) is not. Conclusion about the
differences of inelastic rates for weak localization and conductance
fluctuation is a consequence of missing the diagram (b).} 
\label{Fig5}
\end{figure}

Before we perform actual calculation in Eq.~(\ref{cond4}),
we pause for a moment to discuss a relation of this formula with the
other theoretical work\cite{Blanter}. We observe that the
propagator $\langle{\cal P}_D^{12}\rangle_{\varphi}$ contains all
possible interaction lines drawn between $G^R$ and $G^R$, and also
between $G^A$ and $G^A$, but not between $G^R$ and $G^A$. This is
exactly an object (let us call it {\em CF diffuson}), which determines
the out-scattering term in the collision integral in the Boltzmann
equation, and it was studied in details in
Ref. \onlinecite{Blanter}. In contrast to the ``ordinary'' diffuson,
which is insensitive to electron-electron interaction due to Ward's
identity (charge conservation), the CF diffusion $\langle{\cal
P}_D^{12}\rangle_{\varphi}$ acquires a massive pole, real part of
which is identified with the out-scattering time $\tau_e$. One can
thus imagine (and this was, indeed, conjectured in
Ref.~\onlinecite{Blanter}) that the temperature dependence of
conductance fluctuations is governed by the time $\tau_e$, which is
parametrically different from $\tau_{\phi}$. The calculation presented
below shows that this conjecture is not correct. The resolution of
this fallacy is that the averaging in Eq.~(\ref{cond4}), which is
essentially coupling of all random fields $\varphi^{\alpha}$ according
to the rules (\ref{fdt1}), produces not only a contribution which
contains averages $|\langle{\cal P}^{12}\rangle_{\varphi}|^2$ 
(Fig.~\ref{Fig5}a), but also diagrams where interaction lines connect
upper and lower Green's functions within the same ring
(Fig.~\ref{Fig5}b). Both contributions diverge in the infrared limit
(and have to be regularized in order to extract sensible results
\cite{Blanter}), but their sum is well-behaved.    

To proceed with the evaluation of Eq.~(\ref{cond4}), we write
${\cal P}^{\alpha\beta}$ as a functional integral \cite{Hibbs,AAK}, 
\begin{eqnarray}
& & {\cal P}^{\alpha\beta}_{D,C} (x, x'; t;
\left\{\varphi (x, t)\right\}) = \frac{\theta(t)}{{\cal Z}}
e^{-\frac{t}{\tau_{D,C}^{\alpha\beta}}} \int^{y(t) = x}_{y(0) = x'}
{\cal D}y(\tau)  
\nonumber
\\ 
& & \times \exp \left( \int_0^t d\tau \left\{ -\frac{{\dot y}^2
(\tau)}{4D} + i\varphi^{\alpha} [y(\tau), \tau] - i\varphi^{\beta}
[y(\tau), \tau] \right\} \right), \nonumber\\
 \label{contint1}
\end{eqnarray}
where $\theta(t)$ is the step function,
 ${\cal Z}$ is the normalization factor,
that will be included in the measure of the functional integration
in all of the subsequent formulas. 
Substituting this expression
into Eq.~(\ref{cond4}), 
and averaging over Gaussian random fields ($\langle e^{i\varphi}\rangle
=  e^{-\langle \varphi^2\rangle/2}$), we obtain with the help of
(\ref{fdt1})
\begin{eqnarray}
&& \overline{ \delta G(H_1) \delta G(H_2)} 
=  \frac{(2e^2 D)^2}{3\pi T L^4}\int dx_1 dx_2
\int_0^\infty dt
\label{cond40}
\\ 
&& \times \left(e^{-\frac{2t}{\tau_{D}^{12}}}+
e^{-\frac{2t}{\tau_{C}^{12}}}
\right)
\int^{y_1(t)=x_1}_{y_1(0)=x_2} {\cal D}y_1(t)
\int^{y_2(t)=x_1}_{y_2(0)=x_2} {\cal D}y_2(t) 
 \nonumber  \\
&& \times\exp\left\{ -\int_0^t
dt' \left[ \frac{{\dot y}_1^2}{4D} + \frac{{\dot y}_2^2}{4D}
+\frac{2T}{D\nu} \left\vert y_1(t') - y_2 (t')
\right\vert \right] \right\}.  \nonumber
\end{eqnarray}
Following Ref. \onlinecite{AAK}, we introduce new variables, 
\begin{displaymath}
z_{1,2} (t) = \frac{y_1(t) \pm y_2(t)}{\sqrt{2}}.
\end{displaymath}
This yields
\begin{eqnarray} \label{cond5}
 && \overline{ \delta G(H_1) \delta G(H_2)} 
=  \frac{(2e^2 D)^2}{3\pi T L^4} \int dx_1 dx_2 \int_{0}^\infty dt \\
&& \quad\times
J_1 (\sqrt{2} x_1, \sqrt{2}x_2; t) J_2 (0,0; t)
\left(e^{-\frac{2t}{\tau_{D}^{12}}}+
e^{-\frac{2t}{\tau_{C}^{12}}}
\right)
 \nonumber 
\\ 
&& J_1 (x_1,x_2; t)=\int^{z_1(t)=x_1}_{z_1(0)=x_2} Dz_1(t) 
\exp\left( -\int_0^t
dt'  \frac{{\dot z}_1^2}{4D}\right) 
, \nonumber 
\\ 
&& J_2 (x_1,x_2; t)=\int^{z_2(t)=x_1}_{z_2(0)=x_2} Dz_2(t) 
\nonumber\\
&& \times
\exp\left\{ -\int_0^t
dt' \left[ \frac{{\dot z}_2^2}{4D}
 + \frac{2\sqrt{2}Te^2}{\sigma_1} \vert z_2 (t') \vert \right] \right\} 
, \nonumber 
\end{eqnarray}
where $\sigma_1$ is the one-dimensional conductivity, and we
used Einstein relation $\sigma_1=e^2\nu_1D$. 

Now we represent these functional
integrals $J_{1,2}$ as solutions of differential equations. 
The integral $J_1$ solves 
\begin{equation} \label{int11}
\left( \partial_t - D \partial^2_{x_1} \right) J_1 =
 \delta(t) \delta(x_1 - x_2). 
\end{equation}
Integrating both sides of Eq.~(\ref{int11}) over $x_1$ and $x_2$, and
neglecting the boundary term at $t \ll L^2/D$, we obtain
\begin{equation}
\label{dj1}
\int dx_1 dx_2 J_1 (\sqrt{2}x_1, \sqrt{2}x_2, t) = 
\frac{L}{\sqrt{2}} \theta(t).
\end{equation}
Similarly, $J_2$ obeys the equation
\begin{eqnarray} \label{int12}
\left( \partial_t - D \partial^2_{x_1} +
\frac{2\sqrt{2}Te^2}{\sigma_1} 
\vert x_1
\vert \right)  J_2= \delta(t) \delta(x_1-x_2). 
\end{eqnarray}
Substituting Eq.~(\ref{dj1}) into Eq.~(\ref{cond5}), and using
Eq.~(\ref{int12}), we find
\begin{equation} \label{cond6}
\overline{ \delta G(H_1) \delta G(H_2)}= 
\frac{(2e^2D)^2}{3\sqrt{2}\pi TL^3}
\left(Q_{D}(x=0)+Q_{C}(x=0)\right),
\end{equation}
and $Q_{D,C}(x)$ obeys the equation
\begin{eqnarray} \label{int13}
\left( \frac{2}{\tau^{12}_{C,D}} - D \partial^2_{x} +
\frac{2\sqrt{2}Te^2}{\sigma_1} 
\vert x
\vert \right)  Q_{C,D}(x) =  \delta(x). 
\end{eqnarray}
Equation (\ref{int13}) has been previously considered in
Ref.~\onlinecite{AAK}, and it has the solution  in terms of the Airy
function $Ai(x)$, 
\begin{equation}
 Q_{C,D}(x) = - \frac{L_{\phi}}{2\sqrt{2}D} \frac{Ai\left(
\frac{\tau_\phi}{\tau_{C,D}^{12}}+
\frac{\sqrt{2}|x|}{D L_\phi}\right)}
{Ai^\prime\left(\frac{\tau_\phi}{\tau_{C,D}^{12}}\right)} \ ,
\label{int14}
\end{equation}
where the dephasing time $\tau_\phi$ and the dephasing length
$L_\phi$ have exactly the same form as for the weak localization
correction\cite{AAK,AA} (numerical coefficient is corrected
in Ref.~\onlinecite{AAG}),
\begin{equation}
\frac{1}{\tau_\phi} =
\left(\frac{e^2T\sqrt{D}}{\hbar^2\sigma_1}\right)^{2/3},
\quad L_\phi =\sqrt{D\tau_\phi}\ .
\label{tauphi}
\end{equation}
Substituting Eq.~(\ref{int14}) into Eq.~(\ref{cond6}),
one finally obtains
\begin{eqnarray}
&&\overline{ \delta G(H_1) \delta G(H_2)}=
\left(\frac{e^2}{\hbar}\right)^2
\frac{\hbar D}{3\pi L^2 T}
\frac{L_\phi}{L} \nonumber\\
&&\times
\left[\eta\left(
\frac{\tau_\phi}{\tau_{D}^{12}}\right) 
+\eta\left(
\frac{\tau_\phi}{\tau_{C}^{12}}\right) 
\right], \nonumber \\
&&\eta(x)  = - \frac{1}{[ \ln Ai(x) ]'}. 
\label{final}
\end{eqnarray}

Equation (\ref{final})  with entries (\ref{times}) and (\ref{tauphi})
is the main quantitative result of the present Section. It shows that
the dephasing rate governing temperature and magnetic field dependence
of the mesoscopic fluctuations is {\em exactly} the same as in weak
localization. Moreover, this result can be combined with the
expression for the weak localization correction 
\[
\delta G_{WL}(H_1) = \frac{\delta\sigma_{WL}(H_1)}{L}
;\
\delta \sigma_{WL}(H_1) = -\frac{e^2L_{\phi}}{\pi\hbar}
\eta\left(
\frac{\tau_\phi}{\tau_{C}^{11}}\right)
\]
to the form free of geometrical uncertainties 
(as well as uncertainties in the value of the diffusion coefficient),
\begin{eqnarray}
&&\overline{ \delta G(H_1) \delta G(H_2)}= 
\left(\frac{e^2}{\hbar}\right)
\frac{\hbar D}{3 L^2 T}
\label{final2}\\
&&
\left\vert\delta G_{WL}\left(\frac{H_1-H_2}{2}\right)
+
\delta G_{WL}\left(\frac{H_1+H_2}{2}\right)
\right\vert.
\nonumber
\end{eqnarray}
This result gives the relation between two measurable quantities,
and thus may serve as a test for the dephasing mechanism.
Equations (\ref{final}) and (\ref{final2}) are valid provided
$\hbar/\tau^{12}_{D,C} \ll T$. It is also assumes that there is no
spin-orbit interaction. It may be shown that in the case of strong
spin-orbit (SO) interaction, the result (\ref{final2}) still holds up
to a numerical factor of $1/2$. In the case of the crossover between
strong and weak SO interaction one has to identify the singlet $\delta
G_s$ and triplet  $\delta G_{t}$  contributions to the weak
localization correction $\delta G_{WL} = 3\delta G_{t}-\delta G_{s}$
by corresponding fits and replace $\delta G_{WL}$ in
Eq.~(\ref{final2}) with $\left[\delta G_{WL}+2\delta G_{s}\right]/2 =
\left[3\delta G_{t}+\delta G_{s}\right]/2$.   

Now, for conceptual clarity, we employ the result (\ref{final}) to
extract the relaxation time associated with conductance
fluctuations. It is important that this time is unphysical by itself, 
and only has a meaning when explicitly linked to Eq. (\ref{final}).  

For this purpose, we take $H_1 = H_2 = 0$ and define the time $\tau_T$
as a mass in the
pole in the CF diffuson ${\cal P}_D^{12}$ and CF cooperon ${\cal
P}_C^{12}$ which enter Eq. (\ref{cond3}). Writing 
\begin{displaymath} 
{\cal P}_{C,D}^{12} (x, x', t) = \int \frac{dq d\omega}{(2\pi)^2}
e^{iq(x - x') - i\omega t} \frac{1}{Dq^2 - i\omega +
\tau_T^{-1}}\ ,
\end{displaymath}   
substituting this expression into Eq. (\ref{cond3}) and performing the
integration, we obtain for conductance fluctuations
\begin{equation} \label{cond8}
\overline{\delta G^2} = \frac{(2e^2D)^2}{6\pi\hbar TL^3} \left(
\frac{\tau_T}{D} \right)^{1/2}.
\end{equation}
Comparing this to the result (\ref{final}), we identify the
inelastic relaxation time $\tau_T$ responsible for the temperature
dependence of conductance fluctuations, 
\begin{equation} \label{taut1D}
\tau_T = \eta^2 (0) \tau_{\phi} \approx 0.53 \tau_{\phi},
\end{equation}
where $\tau_\phi$ is defined in Eq.~(\ref{tauphi}), {\em i.e.} it is 
precisely the same time one obtains if one considers weak localization
by introducing a finite mass in the pole of the Cooperon.
Thus, the temperature dependence of conductance fluctuations does not
produce a new time scale as compared to Eq.~~(\ref{tauphi}) 
and is certainly
not determined by the out-scattering time $\tau_e$. The numerical
coefficient $0.53$ reflects the behavior of the scaling function
(\ref{final}) in low magnetic fields.   

\section{Two-dimensional case}
\label{sec:4}

Equation (\ref{final2}) can be readily generalized to the two dimensional
sample, and we outline the main steps of the corresponding derivation.

Consider  a two-dimensional system of the size $L$. Performing the same
steps as in the derivation of Eq. (\ref{cond4})
one finds:
\begin{eqnarray} \label{2cond4}
&& \overline{ \delta G(H_1) \delta G(H_2)} 
=  \frac{(2e^2 D)^2}{3\pi T L^4}\int d^2 \bbox{r}_1 d^2 \bbox{r}_2 
\int dt \\
&&\quad\times 
\left[\langle\left|{\cal P}_D^{12} (\bbox{r}_1, \bbox{r}_2,t)\right|^2
\rangle_{\varphi}
+\langle
\left| {\cal P}_C^{12} (\bbox{r}_1, \bbox{r}_2;
t)\right|^2\rangle_{\varphi} \right],
\nonumber
\end{eqnarray}
where two-dimensional integrations are performed within the sample,
$\langle\dots\rangle_{\varphi}$ stand for the averaging over
the fluctuating field $\varphi^{1,2}$ with correlation function analogous
to Eq.~(\ref{fdt1}),
\begin{eqnarray} \label{2fdt1}  
& &  \langle \varphi^{\alpha} (\bbox{r}, t) \varphi^{\beta}
(\bbox{r}',t') \rangle \nonumber \\
& = & \delta_{\alpha\beta} \delta (t - t') \frac{2T}{D\nu_2} 
\int
\frac{d^2q}{(2\pi)^2} \frac{ e^{i\bbox{q}(\bbox{r}- \bbox{r}')}}{q^2},  
\end{eqnarray}
with $\nu_2$ being the thermodynamic density of states per unit area.
In Eq.~(\ref{2fdt1}), the integration is limited from above 
by $|q| \simeq (T/D)^{1/2}$. Such an accuracy of the ultraviolet
cut-off is sufficient for the logarithmically divergent integral.

Diffuson and cooperon propagators entering Eq.~(\ref{2cond4})
are the solutions of the two-dimensional analog of Eq.~(\ref{difint}), 
\begin{eqnarray} \label{2difint}
 &&\left[  \partial_{t}  - D
\left\{\matrix{\nabla_D^{\alpha\beta}\cr
\nabla_C^{\alpha\beta}
}\right\}^2
 + i \left(\varphi^{\alpha} (\bbox{r}, t) -
\varphi^{\beta} (\bbox{r}, t)\right) 
+ \left\{\matrix{\frac{1}{\tau_D^{\alpha\beta}} \cr
\frac{1}{\tau_C^{\alpha\beta}}
}\right\}\right]\nonumber\\
&& \quad \times 
\left\{\matrix{
{\cal P}^{\alpha\beta}_D (\bbox{r}, \bbox{r}'; t;
\left\{\varphi^{\alpha} (\bbox{r}, t),
\varphi^{\beta}(\bbox{r}, t)
\right\})\cr
{\cal P}^{\alpha\beta}_C(\bbox{r}, \bbox{r}'; t;
\left\{\varphi^{\alpha} (\bbox{r}, t),
\varphi^{\beta}(\bbox{r}, t) \right\}
)}\right\} \nonumber\\
& & \quad \quad = \delta (\bbox{r} - \bbox{r}') \delta (t), 
\end{eqnarray}
where times $1/\tau_{D,C}$, see Eq.~(\ref{times}),
describe the effect of the magnetic field component parallel to the
film plane. The effect of the magnetic field perpendicular to the
plane is described by 
\begin{eqnarray}
&&\bbox{\nabla}_\gamma^{\alpha\beta} \equiv \bbox{\nabla}
+ \frac{ie}{c}\bbox{A}^{\alpha\beta}_\gamma,
\quad \alpha,\beta=1,2; \ \ \gamma = D,C;
\nonumber\\ 
&&\bbox{A}^{\alpha\beta}_D=\bbox{A}^\alpha -\bbox{A}^\beta;
\quad 
\bbox{A}^{\alpha\beta}_C=\bbox{A}^\alpha +\bbox{A}^\beta\ ,
\label{nablas}
\end{eqnarray}
where the vector potentials are such that 
\[
\bbox{\nabla} \times \bbox{A}^\alpha = \bbox{H}^\alpha_\perp \ ,
\]
and $\bbox{H}^\alpha_\perp$ is the component of is the magnetic field
perpendicular to the plane.

Transformations leading to Eqs. (\ref{cond4}) and (\ref{difint}) are
pretty much the same as in 1D provided we make obvious changes $x \to
\bbox{r}$, $q \to \bbox{q}$, $\partial_x \to
\nabla_{\bbox{r}}$. Writing again the CF diffusons and cooperons
${\cal P}_{D,C}^{\alpha\beta}$ as functional integrals
(\ref{contint1}) and performing an averaging over Gaussian fields
$\varphi^{\alpha}$, we obtain a two-dimensional analog of
Eq.~(\ref{cond40}),  
\begin{eqnarray}  
\label{cond21}
&& \overline{ \delta G(H_1) \delta G(H_2)} 
=  \frac{(2e^2 D)^2}{3\pi T L^4}\int d^2\bbox{r}_1 d^2\bbox{r}_2
\int_0^\infty dt \\
&&\times \sum_{\gamma=D,C} e^{-\frac{2 t}{\tau_{\gamma}^{12}}}
\int^{\bbox{y}_1(t)=\bbox{r}_1}_{\bbox{y}_1(0)=\bbox{r}_2} 
{\cal D}\bbox{y}_1(t)
\int^{\bbox{y}_2(t)=\bbox{r}_1}_{\bbox{y}_2(0)={\bbox{r}_2}} {\cal
D}\bbox{y}_2(t) \nonumber \\ 
&& \quad \times \exp\left\{ -\int_0^t 
dt' \left[ \frac{\bbox{\dot y}_1^2}{4D} + \frac{\bbox{\dot y}_2^2}{4D}
\right. \right. \nonumber \\
&& \quad + \left. \left. \frac{ie}{c} 
\left(\bbox{A}_\gamma^{12}(\bbox{y}_1)\bbox{\dot y}_1
- \bbox{A}_\gamma^{12}(\bbox{y}_2) \bbox{\dot y}_2
\right)
\right.\right. \nonumber \\
&& \quad + \left. \left. \frac{4T}{D\nu} \int \frac{d\bbox{q}}{(2\pi)^2}
\frac{1}{q^2} \left[ 1 - \cos(\bbox{q}(\bbox{y_1} (t') -
\bbox{y_2}(t')) \right]\right] \right\}.
\nonumber 
\end{eqnarray}
Introducing new variables 
\[
\bbox{R}(t) =
\frac{\bbox{y}_1 (t)+ \bbox{y}_2 (t)}{2},
\quad
\bbox{r}(t)=\bbox{y}_1 (t) - \bbox{y}_2 (t),
\]
and reducing the functional integrals back to differential equations,
we obtain the result
\begin{eqnarray} \label{cond22}
& & \overline{ \delta G(H_1) \delta G(H_2)} \nonumber \\
& & = \frac{(2e^2D)^2}{3\pi TL^2}
\left[Q_D(|\bbox{r}| = L_T)+Q_C(|\bbox{r}|=L_T)  \right],
\end{eqnarray}
where $Q_{D,C}(\bbox{r})$ obeys the equation
\begin{equation} \label{diff21}
\left[- D \left( \nabla_{D,C}^{12}\right)^2 + U(r) 
+ \frac{1}{\tau_{D,C}^{12}}\right] Q_{C,D}(\bbox{r}) = 
\delta(\bbox{r}),
\end{equation}
and the potential is given by
\begin{equation} \label{pot1}
U(r) = \frac{2 T}{D\nu} \int \frac{d\bbox{q}}{(2\pi)^2}
\frac{ 1 - \cos(\bbox{qr})}{q^2}
\approx  \frac{ T}{\pi D\nu}\ln\left(\frac{L_T + r}{L_T}\right),
\end{equation}
where the last expression and Eq.~(\ref{cond22}) are written with the
logarithmic accuracy and we take into account the high-momentum
cut-off at $q \sim L_T^{-1}$, $L_T = (D/T)^{1/2}$. 

Equations (\ref{cond22}) -- (\ref{pot1}) should be compared with 
the corresponding expression for the weak localization correction
in two dimensions \cite{AAK,AA},
\begin{eqnarray}
& & \delta\sigma(H_1)  = -\frac{e^2}{\pi \hbar}
C(r=l), \label{WL} \\
& & \left[- D \left( \nabla_{C}^{11}\right)^2 + U(r) 
+ \frac{1}{\tau_{C}^{11}}\right] C(\bbox{r}) = 
\delta(\bbox{r}),
\nonumber
\end{eqnarray}
where the logarithmic divergence should be cut at the elastic mean
free path $l$.

Therefore, we conclude that the relation similar to Eq.~(\ref{final2})
should hold,  
\begin{eqnarray}
&&\overline{ \delta G(H=0) \delta G(H=0)}-
\overline{ \delta G(H_1) \delta G(H_2)}
\label{2Dfinal}\\
&&
= 
\left(\frac{e^2}{\hbar}\right)
\frac{\hbar D}{3 L^2 T}\times
\nonumber\\
&&
\left\vert
\delta \sigma_{WL}\left(\frac{H_1-H_2}{2}\right)
+
\delta \sigma_{WL} \left(\frac{H_1+H_2}{2}\right)
- 2 \delta \sigma_{WL} \left(0\right)
\right\vert.
\nonumber
\end{eqnarray}

It is important to emphasize that the relation (\ref{2Dfinal})
holds even before one starts an approximate solution of 
Eq.~(\ref{diff21}). Note however that the result similar to
Eq. (\ref{final2}) does not hold, since both $\overline{ \delta G
\delta G}$ and $\delta \sigma_{WL}$ diverge logarithmically with
different cut-offs. This is why in Eq. (\ref{2Dfinal}) we had to
subtract zero-field contributions, which cancels logarithmic
divergences. 

The effect of the spin orbit interactions on our
final result (\ref{2Dfinal}) is the same as for one-dimensional
geometry, see discussion after Eq. (\ref{final2}).

We write here the explicit expression\cite{AAK,AA} for the weak
localization correction in two dimensions for the reference purpose,
\[
\delta
\sigma_{WL}(H,T)= -\frac{e^2}{2\pi^2\hbar}
\left[
\ln \frac{1}{\tau\Omega_H} - 
\Psi\left(\frac{1}{2} + \frac{1}{\tau^*\Omega_H}\right)
\right]\ ,
\]
where $\Psi(x)$ is the digamma function, $\Omega_H=4eDH_{\perp}/c \hbar$,
and $\tau_*$ is determined by the equation
\[
 \frac{1}{\tau^*}=\frac{1}{\tau_H} + 
\frac{T}{\hbar}\frac{e^2R_\Box}{2\pi\hbar}
\ln \frac{T}{\hbar/\tau^* +\hbar\Omega_H}.
\] 

Similarly to one dimension, we can also extract the inelastic time
$\tau_T$, defined as a pole of CF diffuson in zero magnetic
field. An explicit calculation gives $\tau_T \approx \tau_{\phi}$. 
This relation contains a numerical coefficient of order one, which can
only be determined by going beyond the logarithmic accuracy. We do not
attempt such a calculation in this paper.   

\section{Discussion and conclusions}

Equations (\ref{final2}) and (\ref{2Dfinal}) are the main results of
our paper. They give exact relations which must hold between two
experimentally observable results for the dephasing by the
electron-electron interaction. The only reason for violation of such a
relation is that other channels of dephasing with small frequency
transfer are present.  Thus, the systematic measurements of dependence
of conductance fluctuations on temperature and magnetic field and
comparing it with the weak localization data obtained on the same
sample may give information on the nature of inelastic interactions in
disordered metals.

We are not aware of attempts to make such a comparison between
inelastic times directly.  However, recently Hoadley, McConville, and
Birge (HMB) \cite{McB,Hoadley} presented very careful measurements of the
magnetic field dependence of $1/f$--noise in silver films. A standard
assumption in the theory of $1/f$--noise in metals (for review, see
Ref.~\onlinecite{Feng}) is that it is produced by low-frequency motion
of impurities. Mathematically, the magnitude of $1/f$--noise in such a
model is given by a set of diagrams identical to those for conductance
fluctuations (Figs.~\ref{Fig3}, \ref{Fig5}) with the only difference
that external and internal rings are described by different impurity
configurations \cite{AS,Stone}. As the result the  field dependence
and the temperature dependence of the noise should be given by
the parametric derivative of Eq.~(\ref{2Dfinal}), {\em i.e.} it should
be expressed through the derivatives of the parallel field dependence of
the weak localization. 

HMB compared the timescale defined as a pole in the diffuson (in our
notations, $\tau_T$), with the phase relaxation time $\tau_{\phi}$,
extracted from their own measurements of the weak localization
correction on the same films. Their procedure results in $\tau_T
\simeq \tau_{\phi}/2.6$, which was interpreted to be consistent with
the theory of Ref.~\onlinecite{Blanter}. Our results (\ref{2Dfinal})
contradict that interpretation. 

To our opinion, the only possible reason of this discrepancy is the
electron-electron interaction in the triplet channel which we did not
take into account. This interaction can be singled out in experiments
with the materials with stronger spin-orbit scattering.  Other sources
of $1/f$--noise seem to be excluded, since the functional form of the
experimentally measured by HMB magnetic field dependence perfectly
fits theoretical predictions.  Dephasing on slow moving impurities
itself, see discussion in Sec.~\ref{mf}, would give a temperature
dependence different than that in experiment and may be ruled out. We
believe that the contradiction between the theory and the experiment
revealed in our paper indicates that the quantitative study of
inelastic processes in mesoscopic samples remains an interesting topic
and deserves future investigation.

\section*{Acknowledgments}

One of us (IA) was supported by Packard foundation. We are grateful to
B.L. Altshuler and N.O. Birge for the discussion of the results.

\end{multicols}


\begin{references}
\bibitem{BarangerStone} It was shown by H.~U.~Baranger and A.~D.~Stone,
Phys. Rev. B {\bf 40}, 8169 (1989), that combinations of the type
$G^RG^R$ are canceled in the expression for the total conductance
(but not for the non-local conductivity) before the disorder
averaging.
\end{references}
\end{document}